# Relaxation Dynamics and Genuine Properties of the Solvated Electron in Neutral Water Clusters

Thomas E. Gartmann,¥ Loren Ban,¥ Bruce L. Yoder, Sebastian Hartweg, Egor Chasovskikh, and Ruth Signorell*

*Department of Chemistry and Applied Biosciences, Laboratory of Physical Chemistry, ETH Zürich, Vladimir-Prelog-Weg 2, CH-8093, Zürich, Switzerland*

¥ These authors contributed equally to this work.

* E-mail: rsignorell@ethz.ch

**Abstract:**

We have investigated the solvation dynamics and the genuine binding energy and photoemission anisotropy of the solvated electron in neutral water clusters with a combination of time-resolved photoelectron velocity map imaging and electron scattering simulations. The dynamics was probed with a UV probe pulse following above-band-gap excitation with a EUV pump pulse. The solvation dynamics is completed within about 2 ps. Only a single band is observed in the spectra, with no indication for isomers with distinct binding energies. Data analysis with an electron scattering model reveals a genuine binding energy in the range of 3.55-3.85 eV and a genuine anisotropy parameter in the range of 0.51-0.66 for the ground-state hydrated electron. All these observations coincide with those for liquid bulk, which is rather unexpected for an average cluster size of 300 molecules.

The broad attention that solvated electrons in their ground and excited states have attracted over many decades can be attributed to them being among the simplest quantum solutes as well as to their important role in a wide range of fields. Many studies have focused on excess electrons in water and anion water clusters[1-37] (and refs. therein). For liquid water, the excited state relaxation dynamics of electrons has been investigated over a broad time window from femtoseconds to beyond picoseconds. This has resulted in the picture of an initially delocalized electron that relaxes to a hydrated electron within ~1 ps; subsequent geminate recombination takes place on a much longer time scale[11, 16] (and refs. therein). Typically, multi-photon ionization of neat water using different detection schemes has been employed in these investigations. Recent time-resolved photoelectron studies have revealed a non-adiabatic transition from excited electronic states (p-states) to the ground electronic state (s-state) of the hydrated electron on a sub-100 fs timescale followed by slow (several 100 fs) relaxation in the ground electronic state[9, 10, 38] (and refs. therein). Analogous studies of the relaxation dynamics in clusters - often after an s to p excitation – were conducted for different cluster sizes in photodetachment (anionic cluster) experiments[12, 14, 15, 31, 35] (and refs. therein), and in neutral clusters after excitation by an extreme ultraviolet (EUV) light pulse.[39]

For the electronic ground state (s) several binding motifs (isomers) have been identified in anionic clusters including fully and partially solvated internal states and surface-bound states (refs.[12, 21, 28, 40] and references therein). Distinct experimental vertical binding energies (VBEs) have been observed for the different isomers, with a strong cluster size dependence. The difference in VBE between internal and surface state amounts to about 1 eV for a cluster of 200 molecules. Extrapolation of the cluster data to the infinite bulk results in binding energies of 3.6 eV and 1.6 eV for the fully solvated internal and surface-bound state, respectively, which lie 1.3 eV and 0.3 eV, respectively, below the value for a cluster with 200 molecules. Most experimental studies of the liquid bulk (liquid microjets) find only a single band in the range 3.3-3.6 eV (refs.[2, 8, 37] and references therein). The only exception is an additional surface-bound state at 1.6 eV reported in ref.[5], which has not been reproduced so far by other studies, although a recent EUV study on neutral clusters again speculates about such as surface state.[39] A large number of overlapping bands from different species in this cluster spectrum makes an assignment to different contributions questionable.

In order to characterize the dynamics and different states of the hydrated electron, photoelectron studies record the photoelectron kinetic energies (eKEs) and in some cases the photoelectron angular distributions (PAD) usually characterized by a single anisotropy parameter β. Our recent investigation on hydrated electrons in liquid water microjets has, however, revealed that these measurement quantities are strongly influenced by electron transport scattering in the liquid and thus depend on the photon energy of the ionizing light source.[37] Typically, experimental VBEs vary by about 1 eV depending on the energy of the ionizing photon. Furthermore, the experimental PADs of the hydrated electron in the liquid correspond to an almost isotropic distribution because of electron transport scattering. Proper analysis of the data thus requires the influence of electron scattering to be taken into account.[37, 41-50] We have shown that corrections of the experimental data by means of a detailed electron scattering model make it possible to retrieve genuine (intrinsic) VBEs and βs of 3.7±0.1eV and 0.6±0.2, respectively.[37, 41, 42] In large clusters the electron scattering is substantially different from the bulk (liquid/amorphous solid) as shown in our recent study.[51] While electron scattering cross sections in the liquid and the amorphous solid are virtually identical within experimental uncertainties, those for clusters are significantly larger and lie between the gas[52] and condensed phase values.[37, 41, 42] The reduced dielectric screening in clusters compared with the condensed phase provides a simple explanation for the increased scattering cross sections in the cluster.[51]

The present study investigates the relaxation dynamics in large water clusters with ~300 molecules after above-band-gap excitation by an EUV photon from a high harmonic laser source (pump) and ionization by a UV probe pulse, and the properties of the resulting ground-state solvated electron. It has so far been unknown how the behavior of the solvated electron in large neutral clusters differs from that in large anionic clusters and in liquid bulk. The role that the environment has on the properties of the solvated electron is of general interest, especially regarding confined environments and interfaces.[21] By analogy to anionic clusters, one would expect significant polarization effects shifting the VBEs of clusters relative to the bulk, and possibly also isomers with different VBEs. Surprisingly this is not what we find. By properly accounting for the effects of electron scattering in the clusters, we have evaluated genuine cluster VBEs and β-parameters, which can be directly compared with the corresponding bulk values.

The experimental setup used for the measurements in the present work has been described previously.[51, 53] We used a velocity map imaging (VMI)[54] spectrometer[55-58] to record time-resolved photoelectron images upon photoionization of water clusters using a pump and probe laser excitation scheme. EUV light was produced by high harmonic generation (HHG),[59] as previously described.[51] A home-built, time-preserving monochromator[60, 61] was used to select the 7$^{th}$ harmonic (~10.9 eV) which was used as the pump. As a probe pulse, we used 266 nm light generated by frequency tripling of the 800 nm output from a Ti:Sapphire laser. A beam of neutral water clusters was produced and characterized in the same manner as in previous work.[51] For the experiments presented in this work, the average cluster size was determined to be <n>=300 via the sodium-doping method.[62, 63] See Section S.1 in the Supplementary Information (SI) for details.

The electron scattering simulations have been described previously in detail.[37, 41, 42, 50, 51] The probabilistic scattering model is based on a Monte Carlo solution of the transport equation. The scattering cross sections, angular dependences, and the energy loss functions of all relevant elastic and inelastic phonon, vibron and electronic scattering channels are explicitly taken into account. Simulations are performed for cluster scattering cross sections (models iii and iv),[51] liquid water scattering cross sections (model i)[51] and gas phase scattering cross sections (model ii).[51] While the scattering cross sections used in models i and iii are taken directly from the previous study,[51] the cross sections in model ii and iv were adapted to account for the lower electron kinetic energies in the current study. See Section S.2 in the SI for details.

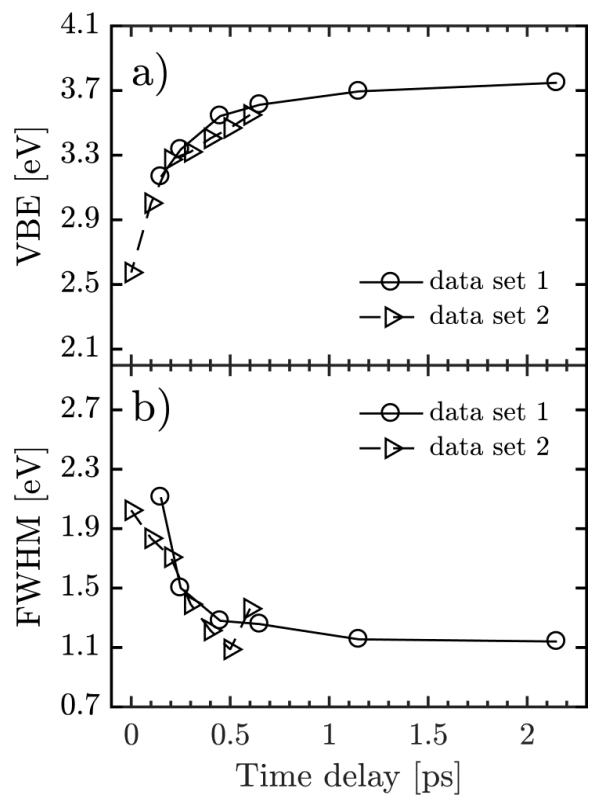

**Figure 1.** Evolution of the photoelectron signal as a function of the pump-probe time delay. a) Vertical electron binding energy (VBE) and b) full width at half maximum (FWHM). Uncertainties are estimated not to exceed ±0.2eV and ±0.3eV for VBE and FWHM, respectively. Lines connecting data points are intended as a guide to the eye. Data sets 1 and 2 represent independent measurement series providing an impression of the reproducibility of the experiment.

Figure 1 shows the time evolution of the vertical electron binding energy (VBE), panel a), and the full-width-at-half maximum (FWHM), panel b), of the binding energy spectra recorded after above-band-gap excitation with EUV light of ~10.9 eV photon energy. The VBE increases with time and stabilizes after 1-2 ps at a value of ~3.75 eV, while the FWHM narrows on the same time scale to stabilize at ~1.14 eV (see Section S.3 in the SI with binding energy spectra and fits at different pump-probe delays in Figure S3). The temporal increase in the VBE reflects the formation of a more strongly bound hydrated electron over time. The fast increase of the VBE by about 1eV in the first few hundred fs cannot be fully resolved in our experiment with a pump-probe cross correlation of about 220fs (FWHM). We assume that after 2 ps a hydrated electron in its electronic ground (s-) state has formed in an equilibrated solvent environment. The direct comparison with previous studies is difficult because the time scales for equilibration depend strongly on how the electron was initially prepared.[32] Still the overall solvation time scale of a few ps is in reasonable agreement with previously reported time-scales for the formation of the ground-state hydrated electron in liquid bulk water and in anionic water clusters[12, 16, 23, 31, 32, 35, 36, 38] (and references therein). Considering that initial ejection lengths for excitation at ~10.9 eV in the liquid are ~3 nm,[11] which exceeds the cluster radius of ~1.3 nm, it is plausible that most of the initial electron density resides near the surface. This situation closely resembles the model calculation of Herbert and coworkers for the localization of surface electrons and their internalization into the bulk.[17] Comparing the calculated time-evolution of the VBE with our experimental results in Fig.1, and accounting for the cross correlation of our experiment we find reasonable agreement with a major change of the VBE in the first 0.5 ps. The initial β-parameter (~0.4) we measure exceeds that of the electronic ground state (~0.19). This might indicate an initially larger s-character or less electron scattering upon ionization. The latter does not appear unreasonable for an electron residing close to the surface.

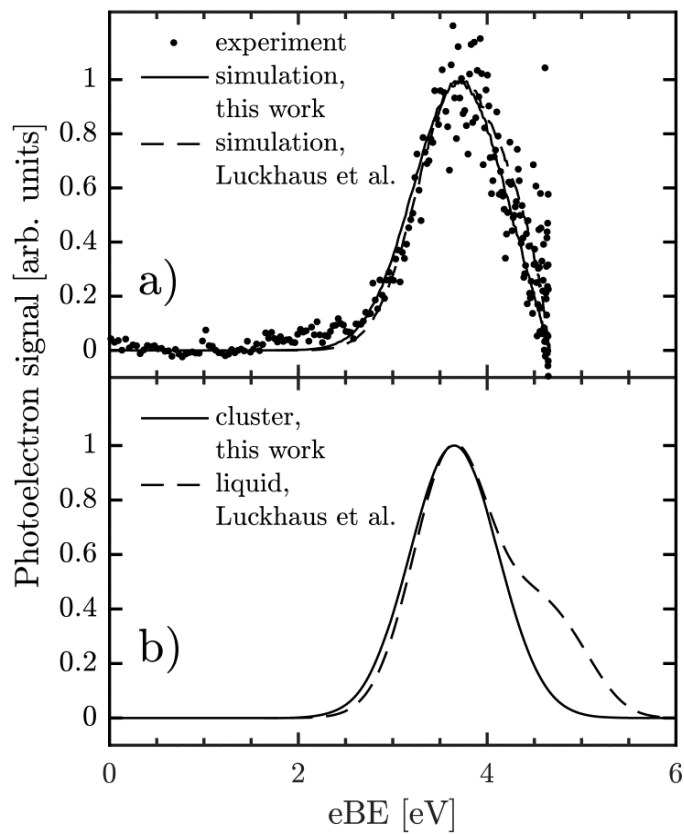

**Figure 2.** a) Experimental binding energy spectrum recorded at a pump-probe delay of ~2 ps (circles). Dashed-line spectrum: Simulation with genuine eBE of the liquid from Luckhaus et al.[37] (dashed line in panel b). Full-line spectrum: Simulation with a single Gaussian fit for the genuine cluster eBE (full line in panel b).

By combining experimental photoelectron VMI images with detailed scattering simulations, we have recently determined genuine vertical binding energies and genuine β-parameters for the ground-state hydrated electron in liquid bulk water.[37] Here, we retrieve genuine VBEs (Figure 2) and genuine β-parameters (Figure 3) for the ground-state hydrated electron in large neutral water clusters from the photoelectron VMI images of the clusters recorded after ~2 ps (Figure 1). Figure 2a shows the experimental binding energy spectrum at 2 ps time delay together with two simulated spectra. For the calculated, dashed-line spectrum we used the genuine binding energy spectrum of the liquid (see starred spectrum in Figure 3 of Luckhaus et al.[37] and dashed-line spectrum in Figure 2b) and the cluster scattering cross sections (model iii) from ref.[51] for the scattering calculations. The calculated, solid-line spectrum is obtained by fitting a genuine binding energy spectrum (solid line in Figure 2b) to the experimental data using scattering calculations with the same cluster scattering cross sections (model iii in ref.[51]). Contrary to the liquid bulk experiments,[37] the experimental cluster data (Figure 2a) are obtained for a single probe energy (266 nm), which is too low to cover the shoulder in the genuine binding energy spectrum of the liquid bulk (dashed-line spectrum in Figure 2b). A single Gaussian is thus sufficient to represent the main part of the genuine binding energy spectrum of the clusters (full line in Figure 2b) covered by the experiment. Both simulations in Figure 2a represent the experimental spectrum very closely, suggesting that the genuine binding energy spectra of the hydrated electron in the liquid bulk and in a neutral water cluster with ~300 molecules are similar. From the scattering correction including a sensitivity analysis using different scattering cross sections for clusters (model iii and iv) from ref.[51] we derive a genuine VBE of 3.55-3.85eV for the hydrated electron in neutral clusters (Table 1). This coincides surprisingly well with the experimental liquid value of 3.7±0.1 eV. As most of the liquid bulk studies, we do not find any evidence for a surface bound electron with a VBE at 1.6 eV (Figs. 2 and S3). Contrary to the anionic clusters, the neutral clusters only show a single VBE band, and even more surprisingly no significant shift relative to the bulk, which one would have expected from polarization effects.

**Table 1**: Values for genuine VBE and genuine β-parameter for the solvated electron in liquid water in the bulk and at the air/water interface region, in neutral water clusters, and in anionic water clusters.

|  | genuine VBE [eV] | genuine β-parameter |
|---|---|---|
| Calc. bulk from refs.[17, 21] | 3.4-3.6 |  |
| Calc. bulk from ref.[64] | 3.75±0.55[(a)] |  |
| Calc. interfacial from refs.[17, 21] | 3.1-3.2 |  |
| Calc. interfacial from ref.[64] | 3.35±0.46[(a)] |  |
| Exp. liquid bulk[37] | 3.7±0.1 | 0.6±0.2 |
| Exp. neutral cluster (300 molecules) this work | 3.55-3.85 | 0.51-0.66 |
| Exp. anion cluster[14] (~50 molecules) |  | ~0.7±0.1 |

[(a)] The quoted uncertainty reflects the widths of the distribution. The uncertainty in the peak positions is probably on the order of ±0.1.

As a result of electron scattering, the PAD measured for bulk liquid water is almost isotropic (β close to 0).[1] Properly accounting for the scattering contributions, this leads to a genuine β-parameter in the liquid bulk of 0.6±0.2.[37] The value seems in fair agreement with what one would expect for an s-like orbital character of the ground-state solvated electron.[21] For clusters with ~300 molecules, we measure a β-parameter of 0.19±0.12 at ~2 ps time delay (dashed horizontal line in Figure 3). (Absolute values for β-parameters are typically not measured to better than ±0.1 accuracy with VMI setups.) The measured value is again strongly reduced compared with the genuine β-parameter in the cluster due to electron scattering processes. To retrieve the genuine β-parameter in the clusters we have performed scattering simulations with the results shown in Figure 3. The diamonds and the squares show the expected measured β-parameter as a function of the genuine β-parameter for scattering simulations with the cluster cross sections from model iii and model iv, respectively. According to the graph the measured β-parameter of 0.19 for the clusters is consistent with a genuine β-parameter in the range 0.51-0.66, depending on the scattering cross sections used for the clusters.[51] As the VBE, the genuine β-parameter of the cluster lies remarkably close to the genuine value of the liquid bulk (Table 1) and to the value of ~0.7±0.1 reported for anionic water clusters containing ~50 molecules.[14] The simulations in Figure S5 in the SI provide an idea of the cluster-size dependence of the measured β-parameter predicted as a function of the genuine β-parameter for clusters with 100 to 900 molecules.

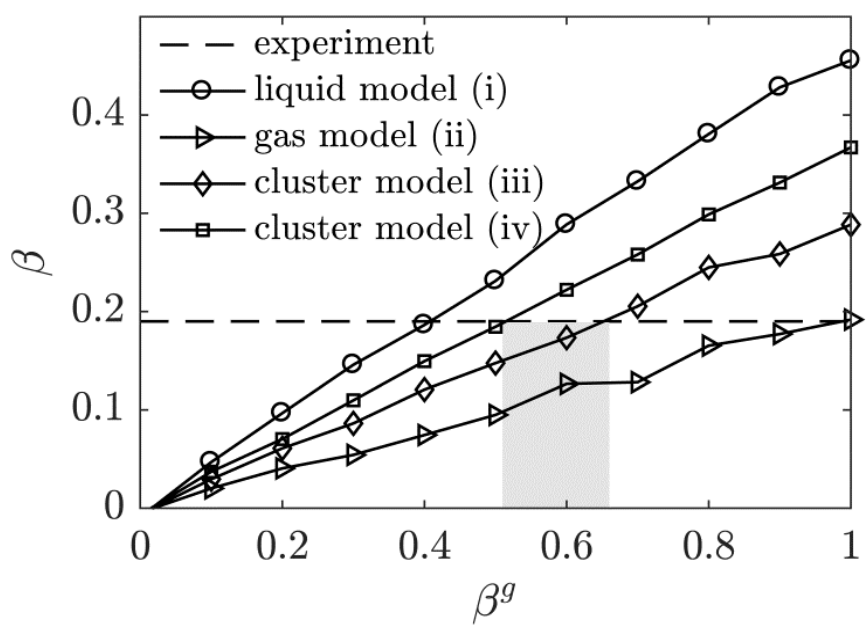

**Figure 3.** Observable β-values as a function of genuine β-values ($\beta^g$) simulated for different models of the electron scattering cross sections (cluster size <n>=300; see ref.[51] for simulation details). Lines in between data points represent linear interpolations. The horizontal dashed line represents the experimental β-value of 0.19. The gray shaded area indicates the range of genuine β-parameters bracketed by cluster models iii and iv, that is consistent with the experimental β-value measured for water clusters.

Figure 3 also shows results for scattering simulations where we have replaced the scattering cross sections of the clusters, either by the scattering cross section of the liquid phase[37, 41, 42] (circles) or by those of the gas phase[52] (triangles). For liquid scattering cross sections, our experimental β-parameter of 0.19 would lead to a genuine β-parameter of ~0.4, while for the gas phase scattering cross sections a genuine β-parameter of ~1.0 would result. Neither would be fully consistent with the results of the liquid bulk. This further underlines the importance of using proper cross sections to account for electron transport in a cluster.

It is rather unexpected to find the same properties for the solvated electron in large neutral clusters as have been found in the liquid bulk, without any need for extrapolation to infinite cluster size. How could this be rationalized? As is the case for most cluster studies, we do not have direct information regarding the temperature and phase of the cluster. For our type of supersonic expansion, we know that we produce warm clusters. It is thus conceivable that we produce liquid-like clusters, i.e. with relatively high mobility of solvent molecules. This is supported by our previous study of the single-photon EUV ionization of pure water clusters generated under precisely the same conditions,[51] with ionization energies extrapolating to the liquid bulk value rather than that of the solid (see Fig. 3 in ref.[51]). The high temperature of our clusters might explain why we do not observe the surface-solvated electron. This is consistent with the depletion of the corresponding signal for anionic clusters in warmer supersonic expansions.[12, 28, 40] However, the fact that the eBE-spectrum both in liquid bulk and in neutral clusters shows only one band does not prove the absence of multiple isomers. It only shows that there are no isomers with significantly different VBEs. Recent calculations by Herbert and coworkers[17, 21, 64] have revealed similar VBE values for electrons solvated in the liquid bulk and at the air-liquid interface region (Table 1). All this still does not explain why the genuine VBE of the liquid bulk is already reached in neutral clusters with 300 molecules without further extrapolation to infinite cluster size, while the VBE of anionic clusters of similar size still lies 1eV below the bulk limit. Similar to our neutral water clusters, the bulk limit of the VBE of the solvated electron in neutral Na-doped water clusters is already reached at very small cluster sizes.[65, 66] Both cases feature a charge separated neutral ground state of the solvated electron and its counter ion ($H^+$(aq) in our case). The presence of the counter ion could explain why the bulk limit is already reached in our neutral clusters, provided that the charge-separated ground state experiences the same polarization shift as the cationic state resulting after ionization.

In conclusion, we have shown that the formation of the ground-state hydrated electron in neutral water clusters with about 300 molecules after above band-gap excitation is completed within about 2 ps. This is roughly consistent with the timescale of the solvation dynamics found in the liquid bulk and in large anion clusters, hinting at similar mechanisms in the different systems. Both the genuine VBE and the genuine β-parameter of hydrated electrons in large neutral clusters virtually coincide with the corresponding liquid bulk values, with only a single band in the spectrum. Contrary to anionic clusters, neither the bulk liquid nor the neutral clusters seem to give rise to different isomers with distinct VBEs. In particular, we do not find evidence for a surface-bound electron with a VBE around 1.6 eV. That the liquid bulk value of the VBE is already reached for clusters containing about 300 molecules is rather surprising and indicates significant polarization effects in the charge-separated neutral ground state. Overall, our results suggest that the nature of the hydrated electron in neutral clusters is similar to that in the liquid bulk. A conclusive proof of this similarity, however, is not possible on the basis of VBEs and β-parameters alone. It will require further investigations with other measurement methods probing different (genuine) properties, and will also need the support of theoretical modelling.

**ACKNOWLEDGEMENTS**
We thank David Stapfer and Markus Steger for technical support. This project has received funding from the European Union's Horizon 2020 research and innovation program from the European Research Council under the Grant Agreement No 786636, and the research was supported by the NCCR MUST, funded by the Swiss National Science Foundation (SNSF), through ETH-FAST, and through SNSF project no. 200020_172472.

**Supporting Information:**

# Relaxation Dynamics and Genuine Properties of the Solvated Electron in Neutral Water Clusters

Thomas E. Gartmann,¥ Loren Ban,¥ Bruce L. Yoder, Sebastian Hartweg, Egor Chasovskikh, and Ruth Signorell

*Department of Chemistry and Applied Biosciences, Laboratory of Physical Chemistry,*

*ETH Zürich, Vladimir-Prelog-Weg 2, CH-8093, Zürich, Switzerland*

¥ These authors contributed equally to this work.

## S.1. Size determination of the neutral cluster distribution:

Figure S1 shows a neutral size distribution of water clusters used in this work. The average cluster size, <n>, was determined to be 300 water molecules via the sodium-doping method[1-3] and analysis as described in detail previously.[4] Panel a) shows a typical mass spectrum of the sodium-doped cluster size distribution. The resulting cluster abundances are shown in panel b) after all relevant corrections were applied.

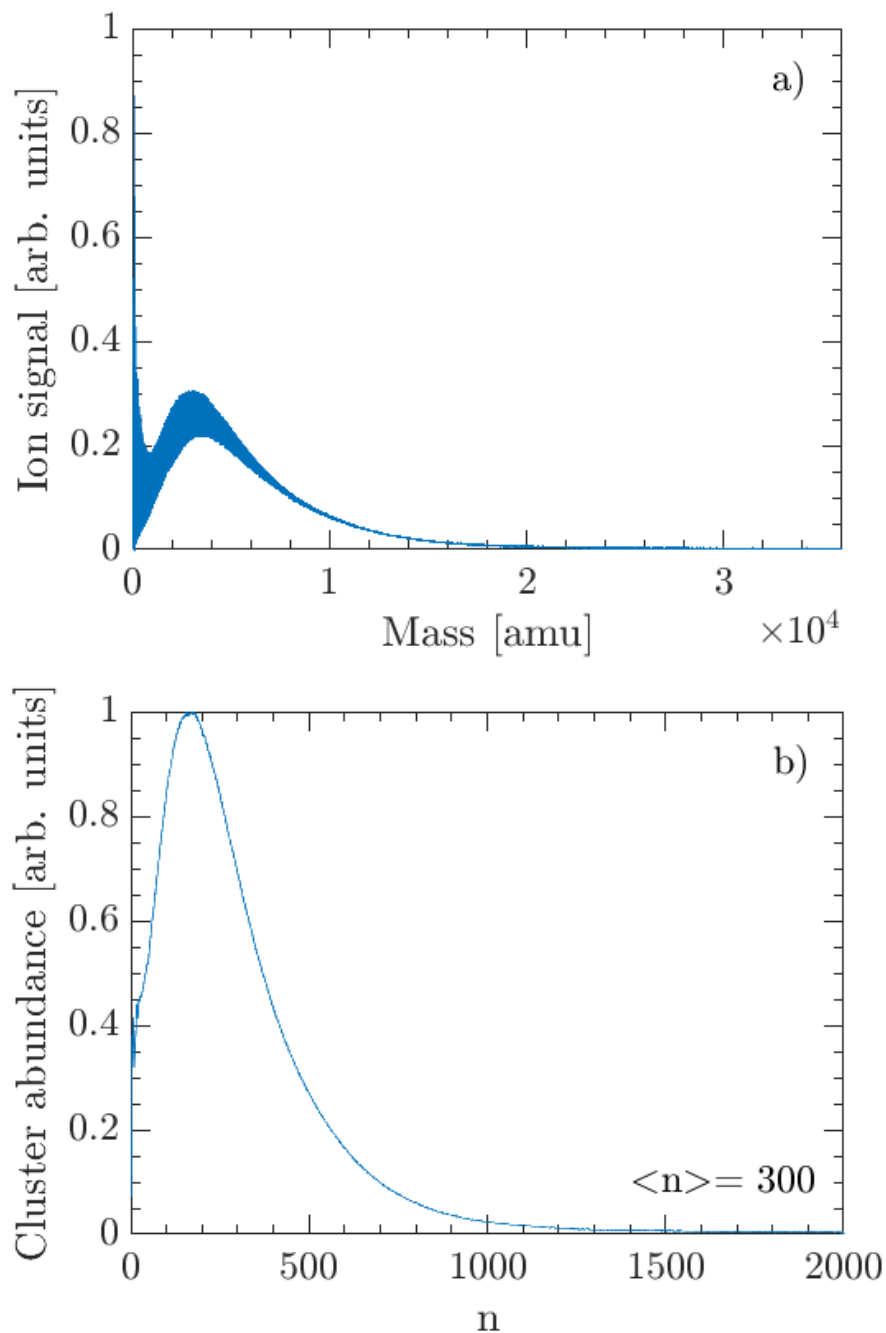

**Figure S1:** a) Typical mass spectrum for cluster size determination, recorded with the Na-doping method. b) Corresponding neutral cluster size distribution with <n> = 300.

## S.2. Determination of electron scattering cross sections for clusters:

To mimic gas phase electron scattering (Figure S2, model (ii), dashed red lines) the liquid bulk scattering cross sections of model (i) were scaled with constant factors to match the corresponding gas phase data recommended by Itikawa and Mason[6] (green circles and green line). The scaling factors for quasi-elastic and vibrational scattering channels are 8 and 7 respectively. Note that electronic scattering channels do not contribute in the energy range below 5 eV kinetic energy for which these scaling factors are determined. The scaling factors used in this study differ from the ones previously determined[4] due to the different electron kinetic energy range considered. The procedure itself however, is identical to previous work.[4] For the quasi-elastic processes (Figure S2 a)), the sum of the liquid bulk elastic scattering cross sections and the isotropic parts of all liquid bulk phonon related scattering cross sections was scaled to match the gas phase momentum transfer cross sections given by Itikawa and Mason.[6] For the vibrational scattering processes (Figure S2 b)) their sum was compared with the sum of all gas phase vibrational cross sections.[6] In order to sum the vibrational scattering cross sections for the gas phase, the data given by Itikawa and Mason was first linearly interpolated.

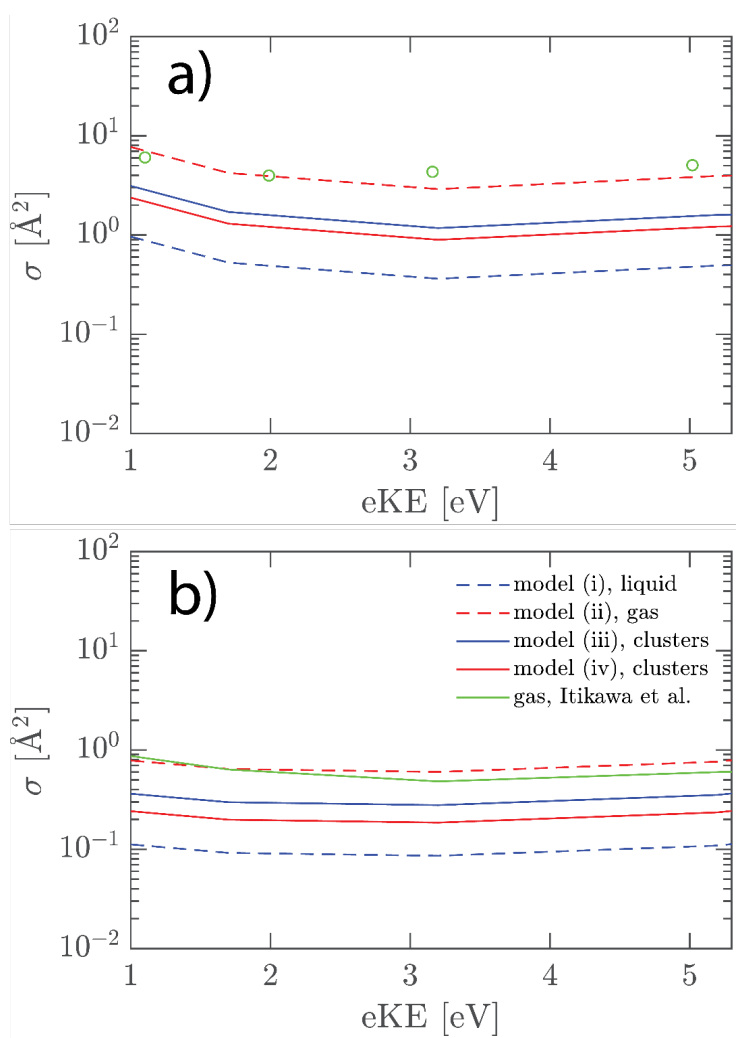

**Figure S2:** Total electron scattering cross sections for quasi-elastic (a) and vibrational scattering (b) in liquid water (blue dashed lines), gas phase water (red dashed lines) and clusters (blue and red full lines). Gas phase scattering cross sections derived from Itikawa and Mason[6] are shown as green circles and full green lines.

To obtain scattering cross sections that reproduce the situation in the clusters (model (iii) and model (iv)) the liquid and the gas phase scattering cross sections are scaled with the square of the dielectric constant at optical frequencies ($1.8^2=3.24$). This scaling factor is used to multiply the scattering cross sections of the liquid bulk (model (i)) to describe a situation without dielectric screening (model (iii)), as an intermediate case between the gas phase and the liquid phase. An alternative intermediate case (model (iv)) is created by dividing the gas phase scattering cross sections (model (ii)) by 3.24. Both models ((iii) and (iv)) have been shown to qualitatively describe electron scattering in neutral water clusters with < 1000 molecules.[4]

The idea behind the approach is that in the clusters where most scattering events take place within a few molecular layers (on the order of 5 or less) away from the surface, the dielectric screening is strongly reduced compared with the bulk. On the timescale of individual scattering events (sub-femtosecond) only the electronic contribution to the dielectric constant, i. e. the optical dielectric constant, is important. Its value is assumed to be constant throughout the cluster. Note that the large reduction of the dielectric constant at water interfaces discussed in the literature does not concern the electronic contribution. Since the scattering cross sections scale with the square of the interaction between the electron and the scattering center, the dielectric constant enters quadratically.

## S.3. VMI data analysis:

Recorded velocity map images were reconstructed using MEVELER[5] resulting in photoelectron kinetic energy and angular distributions, described with a single anisotropy parameter $\beta$

$$I(\theta) \propto 1 + \frac{\beta}{2}(3\cos^2\theta - 1),$$

where $I(\theta)$ is the signal at angle $\theta$ defined between the laser polarization vector and the ejection direction of the photoelectrons. Photoelectron spectra were fitted using a single Gaussian function to extract the vertical binding energies (VBEs) and full widths at half maximum (FWHMs) as a function of the pump-probe delay (Figure S3). The fit was performed for the data points in the eBE region below ~4.1 eV. The photoelectron anisotropy parameter $\beta$ was subsequently determined as an average over the half width at half maximum obtained from the fit.

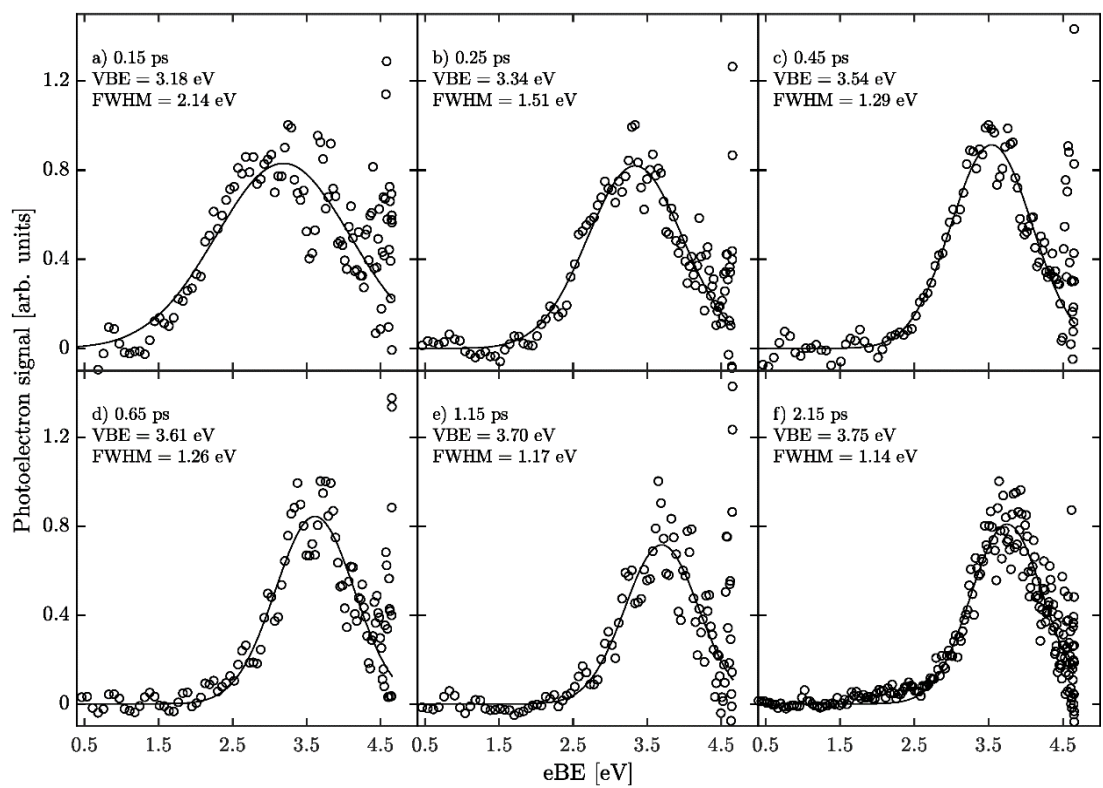

**Figure S3:** Time-dependent photoelectron spectra and corresponding Gaussian fits used to determine VBEs and FWHMs. Spectra for delays <2 ps have been smoothed by a 3 point moving average.

## S.4. Size-dependence of the VBE and the β-parameter:

Figure S4 shows simulated photoelectron spectra as a function of cluster size given as number of water molecules, n. The simulations used the same genuine parameters that were determined from the fit to the experimental results (Figure 2 in the main text). The cluster size has a very minor influence on the kinetic energy spectra in the range n=100-900 water molecules. By contrast, the effect of cluster size is more pronounced for the $\beta$-parameter (Figure S5).

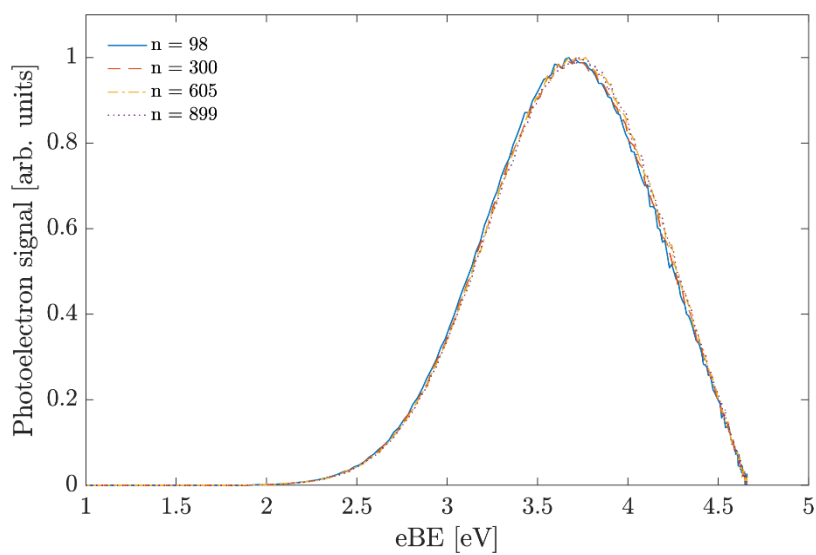

**Figure S4:** Size-dependence of the simulated photoelectron spectra. Simulations were performed with cluster model iii.

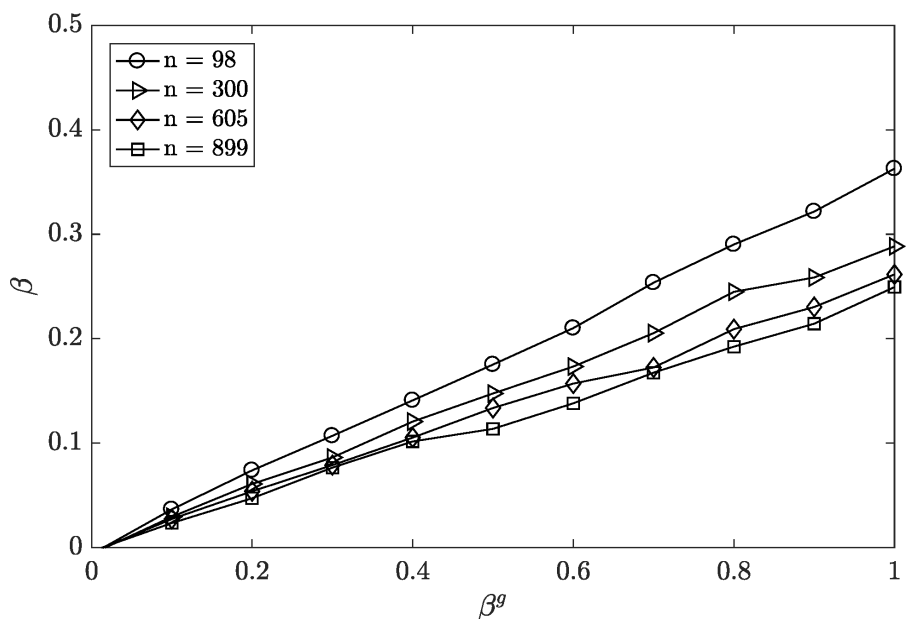

**Figure S5:** Size-dependence of the observable $\beta$ value as a function of the genuine $\beta$-parameter ($\beta^g$). Simulations were performed with cluster model iii. Lines in between data points represent linear interpolations.